\documentclass[nofootinbib,preprint]{revtex4-1}
\usepackage{amsmath}
\usepackage{amsthm}
\usepackage[colorlinks=true,urlcolor=blue,citecolor=blue,linkcolor=blue]{hyperref}
\usepackage[shortlabels]{enumitem}
\usepackage{amsfonts}
\usepackage[caption = false]{subfig}
\usepackage[final]{graphicx}
\usepackage{multirow}
\usepackage{float}
\usepackage{bbold}
\usepackage{stmaryrd}
\usepackage{color}
\usepackage{hyperref}
\usepackage{verbatim}
\usepackage{cases}
\usepackage{amsfonts}
\usepackage{amssymb}
\usepackage[]{mathrsfs}
\usepackage[table]{xcolor}
\usepackage{epsfig}
\usepackage{bm}
\usepackage{setspace}
\usepackage{enumerate}
\usepackage{dcolumn}% Align table columns on decimal point
\usepackage{bm}% bold math
\usepackage{enumitem}
\usepackage{multirow}
\usepackage{multirow, array} % para las tablas
\usepackage{graphicx}% Include figure files
\usepackage{dcolumn}% Align table columns on decimal point
\usepackage{bm}% bold math
\usepackage{color}
\usepackage{amssymb}
\usepackage{lipsum}
\usepackage{amsmath}
\usepackage{graphicx}
\usepackage{array}
\usepackage{multirow}
\usepackage{booktabs}

\usepackage[english]{babel}
\begin{document}

%\preprint{APS/123-QED}

\title{Hydrogen atom confined inside an inverted-Gaussian potential}% Force line breaks with \\
%\footnote{explicar en nota a pie de pagina USMER y sector}
%\author{Author}

\author{H. Olivares-Pilón}%
%\email{alfred.uren@correo.nucleares.unam.mx}
\affiliation{Departamento de F\'{i}sica, Universidad Aut\'onoma Metropolitana Unidad Iztapalapa, San Rafael Atlixco 186, 09340 Cd. Mx., M\'exico}

\author{A. M. Escobar-Ruiz}%
\affiliation{Departamento de F\'{i}sica, Universidad Aut\'onoma Metropolitana Unidad Iztapalapa, San Rafael Atlixco 186, 09340 Cd. Mx., M\'exico}

\author{M. A. Quiroz-Ju\'{a}rez}%
%\email{alfred.uren@correo.nucleares.unam.mx}
\affiliation{Centro de F\'isica Aplicada y Tecnolog\'ia Avanzada, Universidad Nacional Aut\'onoma de M\'exico, Boulevard Juriquilla 3001, Juriquilla, 76230 Quer\'etaro, M\'exico}

%\author{S. A. Cruz}%
%\email{alfred.uren@correo.nucleares.unam.mx}
%\affiliation{Departamento de F\'{i}sica, Universidad Aut\'onoma Metropolitana Unidad Iztapalapa, San Rafael Atlixco 186, 09340 Cd. Mx., M\'exico}

\author{N. Aquino}
\email{naa@xanum.uam.mx}
\affiliation{Departamento de F\'{i}sica, Universidad Aut\'onoma Metropolitana Unidad Iztapalapa, San Rafael Atlixco 186, 09340 Cd. Mx., M\'exico}


\date{August 16, 2022}% It is always \today, today,

\begin{abstract}
In this work, we consider the hydrogen atom confined inside a penetrable spherical potential. The confining potential is described by an inverted-Gaussian function of depth $\omega_0$, width $\sigma$ and centered at $r_c$. In particular, this model has been used to study atoms inside a $C_{60}$ fullerene. For the lowest values of angular momentum $l=0,1,2$, the spectra of the system as a function of the parameters ($\omega_0,\sigma,r_c$) is calculated using three distinct numerical methods: (\textit{i}) Lagrange-mesh method, (\textit{ii}) fourth order finite differences and (\textit{iii}) the finite element method. Concrete results with not less than 11 significant figures are displayed. Also, within the Lagrange-mesh approach the corresponding eigenfunctions and the expectation value of $r$ for the first six states of $s, p$ and $d$ symmetries, respectively, are presented. Our accurate energies are taken as initial data to train an artificial neural network as well. It generates an efficient numerical interpolation. The present numerical results improve and extend those reported in the literature.
\end{abstract}

\maketitle

%\tableofcontents

%\section{\label{sec:level1}First-level heading:\protect\\ The line
%break was forced \lowercase{via} \textbackslash\textbackslash}
%\section{\label{sec:level1}Introducción}

\section{Introduction} 
In the majority of problems the time-independent Schr\"{o}dinger equation does not admit exact solutions in terms of elementary functions or in terms of special functions. In such cases it is necessary to solve this equation by means of numerical and approximate methods which inherently carry a certain degree of accuracy. Some of the most common numerical methods are the following: the Numerov method, the spline-based method, the finite difference, the finite element, the Lagrange mesh method and the variational one, among others.

These numerical schemes are widely used in the study of both spatially confined and unconfined quantum systems. In particular, the investigation of spatially confined quantum systems has gained much interest because some of their physical properties change abruptly with the size of the confining barrier. Furthermore, many physical phenomena can be modeled by a confined quantum system as for example: atoms and molecules subject
to high external pressures, atoms and molecules in fullerenes, inside
cavities such as zeolite molecular sieves or in solvent environments, the specific
heat of a crystalline solid under high pressure, etc. A complete list of
applications can be found in the articles and reviews on the subject \cite{Michel} - \cite{Ley}.

Eighty-five years ago, Michels et. al. \cite{Michel} calculated the variation of the polarizability of hydrogen under high pressure. To this end, they proposed a simple model of the confined hydrogen atom where, to a first approximation, the nucleus is anchored in the center of a spherical box of radius $r_0$ and impenetrable walls. It is assumed that this infinite potential is due to the presence of neighboring negative electric charges. The corresponding wave function vanishes at the surface of the sphere, i. e. it must obey Dirichlet boundary conditions. Since then, this model has been successfully applied in the study of confined many-electron atoms and molecules. However, it takes into account the effects of repulsive forces only. A more realistic potential that embodies an attractive force considers a softer confinement in cavities of penetrable walls. The simplest penetrable confining potential of this type is that of a step potential \cite{Ley2}-\cite{Aquino3}. Some other relevant penetrable confining potentials are the logistic potential \cite{Aquino4} and the inverted Gaussian function \cite{Xie}-\cite{Adam}. Moreover, there are several potentials that have been employed to describe atoms inside fullerenes: the attractive spherical Shell \cite{Dolmatov}, the $\delta$-potential \cite{Amusia} and a Gaussian spherical Shell \cite{NPGM:2011}.

\bigskip 

In this work we calculate the energies and wave functions of the lowest states of a hydrogen atom confined inside a spherical penetrable box. This system has been analyzed in \cite{NPGM:2011} (and references therein). Here, a more accurate and systematic study is carried out. The corresponding confining potential is described by an inverted Gaussian function. To solve the Schr\"{o}dinger equation we employ three different numerical methods: the Lagrange mesh, finite difference and
finite element. They are complemented by the use of an artificial neural network. Making a comparison of the so obtained results we discuss the advantages of each individual method.

\section{Methodology}

The Schr\"odinger equation of hydrogen atom confined in a Gaussian spherical 
shell $w(r)$, in atomic units $(\hbar=m_e=e=1)$,  is of the form
\begin{equation}
\bigg[\,-\frac{1}{2} \Delta  \ - \ \frac{1}{r} \  + \  w(r)\,\bigg]\,\psi\ = \ E\, \psi \ ,
\label{gaussV}
\end{equation}
where $\Delta$ is the 3-dimensional Laplacian, and the Gaussian spherical barrier $w(r)$ is given by
\begin{equation}
w(r)\ = \ -\,\omega_0\, \textrm{exp}[-(r-r_c)^2/\sigma^2]\,,
\label{Gpot}
\end{equation}
here, $\omega_0$ is the well depth, $r_c$ is the position of the center of the peak 
and $\sigma$ is the width of the Gaussian, respectively. In concrete calculations, for the purposes of comparison with the existing results in the literature, from time to time the parameters $r_c$ and $\sigma$ will be presented in Angstroms (\r{A}).

\begin{figure}[t]
\includegraphics[width=9.0cm]{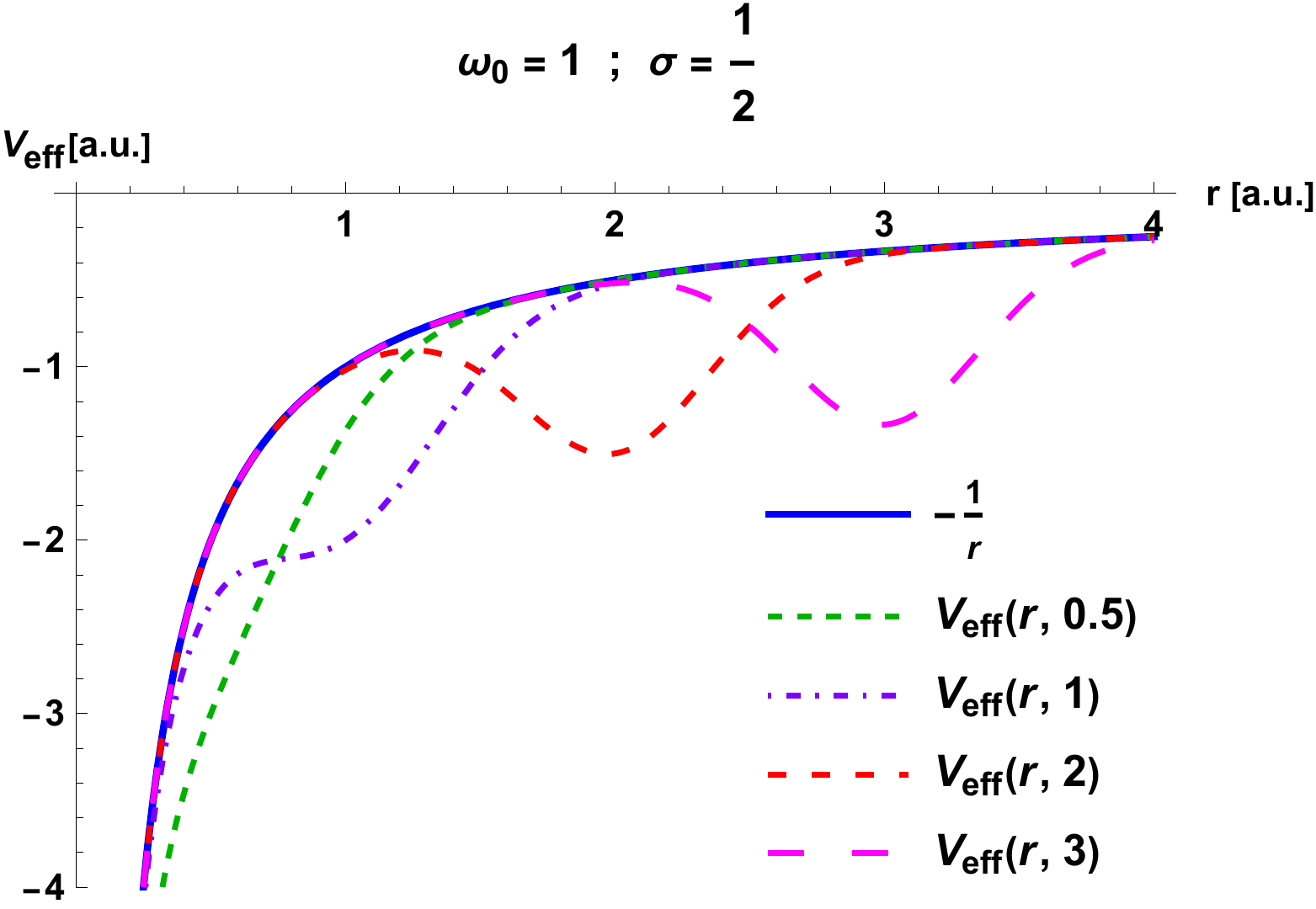}
\caption{Effective potential $V_{\rm eff}(r,\,r_c)$ (\ref{Veffp}) at $l=0$ for $\omega_0=1$ a.u.,
$\sigma=0.5$ a.u. and four different values in a.u. of $r_c=0.5, 1, 2$ and $3$.}
\label{fpotVeff0}
\end{figure}

\begin{figure}[ht]
\includegraphics[scale=0.28]{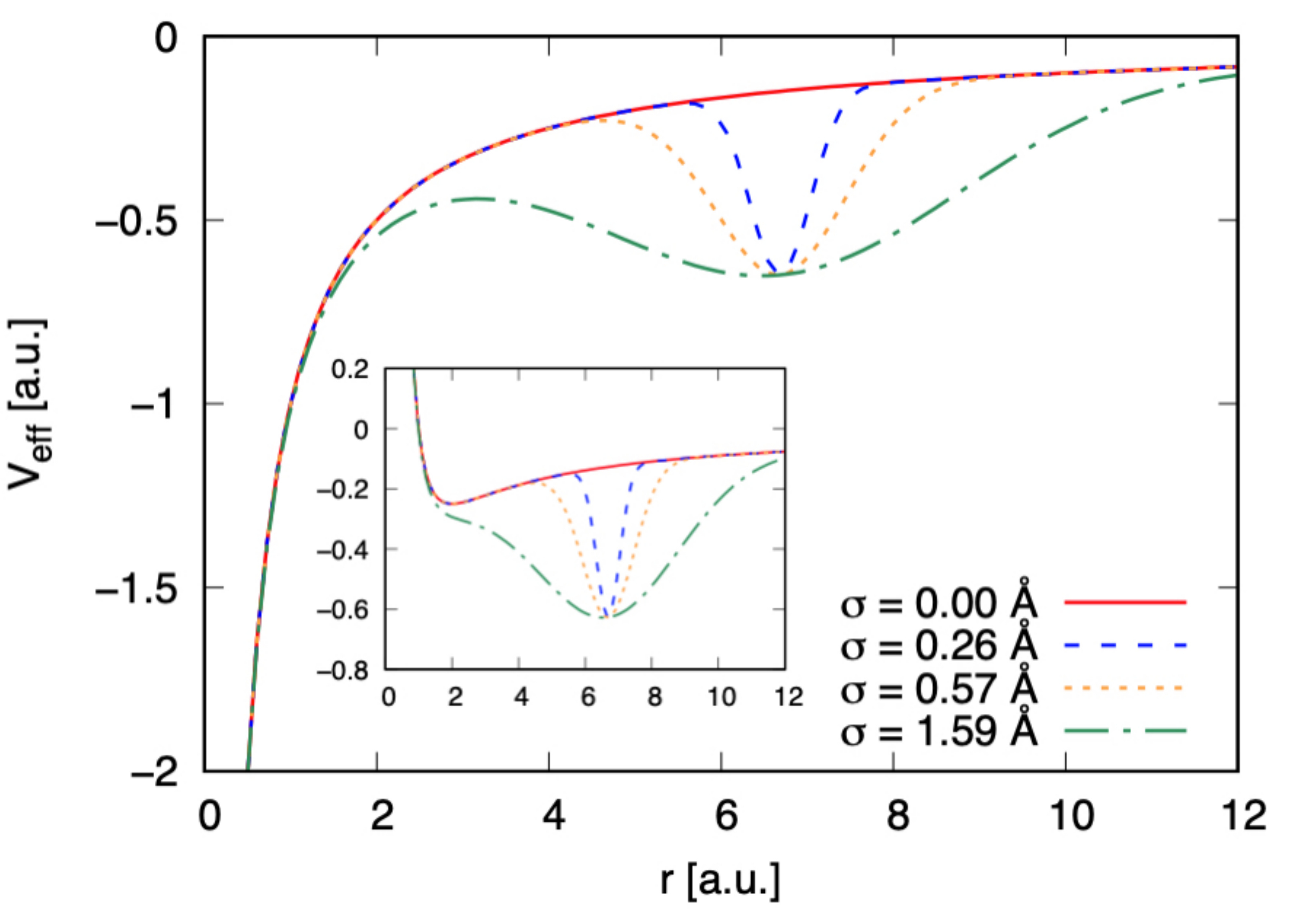}
\caption{Effective potential $V_{\rm eff}(r)$ (\ref{Veffp}) with $\omega_0=0.5$~a.u. 
and $r_c=3.54$~\r{A} for different $\sigma$ values: $\sigma=0$ (continuous red line), 
$\sigma=0.25$~\r{A} (blue dashed line), $\sigma=0.57$~\r{A} (orange dotted line) and 
$\sigma=1.59$~\r{A} (green dash-dotted line). The main figure corresponds to the case 
of $l=0$ while $l=1$ is shown in the inset.}
\label{fpotVeff}
\end{figure}

As it is usual for any
central potential, the angular momentum is conserved and the solutions of the Schrödinger equation (\ref{gaussV}) in spherical coordinates ($r,\theta,\phi$) can be factorized, explicitly
\begin{equation}
\label{factorpsi}
\psi_{nlm}(r,\theta,\phi) \ = \ R_{nl}(r) \,Y_{l,m}(\theta,\phi) \ ,
\end{equation}
with $Y_{l,m}(\theta,\phi)$ being a spherical harmonic function. 
%The radial wave function 
%$R_{nl}(r)$ should satisfy the following differential equation
It is convenient to introduce an auxiliary function $u_{nl}(r)$ defined by the relation $ R_{nl}(r)\ =  u_{nl}(r)/ r $. In this case, from (\ref{gaussV}) and (\ref{factorpsi}) it follows that $u_{nl}(r)$ must obey the isospectral radial problem
\begin{equation}
\bigg[\,-\frac{1}{2}\frac{d^2}{d r^2} \ + \ V_{\rm eff}(r) \,\bigg] u_{nl}(r)\ = \ E\, u_{nl}(r)\ ,
\label{scheq}
\end{equation}
with the boundary condition $u_{nl}(r=0) = 0$.
The effective potential appearing in (\ref{scheq}) is given by
\begin{equation}
\label{Veffp}
V_{\rm eff}(r)\ =\ -\frac{1}{r}\ + \  \frac{l(l+1)}{2\,r^{2}}\ + \ w(r)\ ,
\end{equation}
$l=0,1,2,\ldots,$ is the quantum number of angular momentum.
Formally, (\ref{scheq}) describes a one-dimensional particle of unit mass in the half 
positive line with an effective potential $V_{\rm eff}(r)$ which is displayed in 
Figure~\ref{fpotVeff0}  for fixed $l=0$, $\omega_0=1$ a.u., $\sigma=0.5$~a.u. and four 
different values of $r_c$ (the position of the center of the peak). 
In turn, Figure~\ref{fpotVeff} shows the behavior of $V_{\rm eff}$ (\ref{Veffp}) as a function of 
the width  $\sigma$ of the Gaussian for fixed $\omega=0.5$~a.u. and $r_c=3.54$~\r{A}.    

%%\begin{equation}
%%\label{radialeq}
%%-\frac{1}{2} \frac{1}{r^2} \frac{d}{dr} \left(  r^2 \frac{dR_{nl}}{dr} \right)+ \left[  \frac{l(l+1)}{2 r^2} -\frac{1}{r}+w(r) %%\right] R_{nl}=E\, R_{nl} \ .
%%\end{equation}

%%%\begin{equation}
%%%\label{radialeq}
%%%-\frac{1}{2} \frac{1}{r^2} \frac{d}{dr} \left(  r^2 \frac{dR_{nl}}{dr} \right)+ V_{\rm eff}(r)\, R_{nl}=E\, R_{nl}\,,
%%%\end{equation}
%%%with the effective potential %%%given by:
%%%\begin{equation}
%%%\label{Veffp}
%%%V_{\rm eff}(r)= -\frac{1}{r}+ %%%\frac{l(l+1)}{2r^{2}}+w(r)\,.
%%%\end{equation}

%%%Defining the auxiliary function $u_{nl}(r)$ as
%%%\begin{equation}
%%%R_{nl}(r)\ = \ \frac{ %%%u_{nl}(r)}{r} \ ,
%%%\end{equation}
%%%which satisfies $u_{nl}(r=0) = 0$, the radial Schr\"odinger equation can be written as:
%%%\begin{equation}
%%%\left(-\frac{1}{2}\frac{d^2}{d r^2} +V_{\rm eff}(r) \right) u_{nl}(r)\ = \ E\, u_{nl}(r)\,.
%%%\label{scheq}
%%%\end{equation}

%%\begin{equation}
%%\left(-\frac{1}{2}\frac{d^2}{d r^2} +\frac{l(l+1)}{2\,r^2}-\frac{1}{r} + w(r)\right) u_{nl}(r)\ = \ E\, u_{nl} \ .
%%\label{scheq}
%%\end{equation}
%%The function $u_{nl}$ satisfies
%%\begin{equation}
%%u_{nl}(r=0) \ = \ 0 \ .
%%\end{equation}

In order to solve the radial Schr\"odinger equation~(\ref{scheq}), three different 
methods will be employed: $i$) The Lagrange-mesh method, $ii$) Finite difference 
method, and $iii$) Finite element method. By combining our accurate results with an artificial neural network, we construct an efficient numerical interpolation for the energies.

\subsection{The Lagrange-Mesh Method}

%In order to solve the radial Schr\"odinger equation (\ref{scheq}), the Lagrange-mesh 
%method (LMM) is  used~\cite{BH:1986,DB:2015}.
In the  context of the Lagrange-mesh method (LMM)~\cite{BH:1986,DB:2015},  
a set of $N$ Lagrange functions $f_i\,(x)$ defined 
over the  domain of the radial variable  is associated with $N$ mesh points $x_i$ which 
correspond to the zeros  of Laguerre polynomials of  degree 
$N$, {\it i.e.} $L_N(x_i) = 0$.
The Lagrange-Laguerre functions $f_i\,(x)$ which satisfy the Lagrange
conditions 
\begin{equation}
\label{lagcond}
f_i(x_j)=\lambda_i^{-1/2}\delta_{ij}\,,
\end{equation}
at the $N$ mesh points  are given by
\begin{equation}
\label{lagF}
f_i(x)=(-1)^{i} \frac{x}{x_i^{1/2}}\frac{L_N(x)}{(x-x_i)}e^{-x/2}\,.
\end{equation}
The coefficients $\lambda_i$ are the weights associated with a Gauss quadrature 
\begin{equation}
\label{gqcond}
\int_0^{\infty} G(x) dx \approx \sum_{k=1}^{N} \lambda_k\, G(x_k)\,.
\end{equation}
In terms of the $N$  Lagrange functions $f_i(x)$~(\ref{lagF}), the solution of the Schr\"odinger 
equation~(\ref{scheq}) is expressed  as
\begin{equation}
\psi(r) =\sum_{i=1}^{N}c_i f_i(r)\,.
\label{lmfunc}
\end{equation}

The trial function~(\ref{lmfunc}), together with the Gauss quadrature~(\ref{gqcond}) and the 
Lagrange conditions~(\ref{lagcond})  leads to the system of variational equations
\begin{equation}
\label{lmmeqs}
\sum_{j=1}^N \left[h^{-2}\,T_{ij}+\left(\frac{l(l+1)}{2\, h^2\,x_i^2}-\frac{1}{h\,x_i}+\omega(h\,x_i)\right)\delta_{ij}\right]\,c_j = E\,c_i\,,
\end{equation}
where $\omega(x_i)$ is the Gaussian potential~(\ref{Gpot}) evaluated at the mesh points $x_i$
and $T_{ij}$ are the kinetic-energy matrix elements 
whose explicit expression is found in~\cite{DB:2015}. $h$ is a scaling factor that allows 
to adjust the mesh to the system in consideration.
By solving the system~(\ref{lmmeqs}), not only the energies $E$ are obtained but also the 
eigenvectors $c_{i}$ from which the approximation to the wave function~(\ref{lmfunc}) is obtained.

Inside of the LMM approach, the expectation value of the radial coordinate  $r$ is
easily calculated. Given the approximation to the wave function~(\ref{lmfunc}) together 
with Gauss quadrature~(\ref{gqcond}) and the Lagrange condition~(\ref{lagcond}) leads to  
\begin{equation}
\langle r \rangle = \sum_{i=1}^{N} c_i^2\, x_i\,, 
\end{equation}
where $c_i$ are the  eigenvectors resulting from solving~(\ref{lmfunc}) and 
$x_i$ are the mesh points.

%------------------------------------------------------------------------
%------------------------------------------------------------------------
\subsection{Finite difference method}
The finite difference method (FDM) is a numerical method easy to implement on a
computer, and it is used to solve ordinary and partial differential
equations in an approximate way. The method is based on the discretization
of the Hamiltonian on a spatial grid, replacing the values of the function
and its derivatives by their values at discrete points. 
The Schr\"odinger equation is then solved on a uniform grid~\cite{Kobus}
defined by the set of discrete points which are the nodal points $\left\{ r_{j}\right\}$.

%In present work we
%obtained the solution of the Schr\"{o}dinger equation on a uniform grid \cite{Kobus},
%this set of discrete points are the nodal points $\left\{ r_{j}\right\} $.

%%****
In order to use the finite difference method and the finite element 
method, it is convenient to encase the system inside a spherical box 
with impenetrable walls of radius $r_{\max }$. In this case, the 
Schr\"{o}dinger equation to be solved can be written as (cf.~(\ref{scheq})):
\begin{equation}
-\frac{1}{2}\frac{d^{2}\tilde{u}_{nl}}{dr^{2}}\,+\,V_{\rm eff}(r) \,\tilde{u}_{nl}\,+\,V_{c}\,\tilde{u}_{nl}\,=\,E_{\max}\,\tilde{u}_{nl},
\end{equation}
%%\begin{equation}
%%-\frac{1}{2}\frac{d^{2}\tilde{u}_{nl}}{dr^{2}}\,+\,\left(\frac{l(l+1)}{2r^{2}}\,-\,\frac{\tilde{u}}{r}\,+\,w(r) \right) %%\tilde{u}_{nl}\,+\,V_{c}\,\tilde{u}_{nl}\,=\,E_{\max}\,\tilde{u}_{nl},
%%\end{equation}
where $V_{\rm eff}(r)$ is the effective potential~(\ref{Veffp}) and the 
confinement potential $V_{c}$ is defined as follows
\begin{equation}
V_{c}(r)=\left\{ 
\begin{tabular}{l}
0\ , \qquad \, if $\ r \leq r_{\max }$ \\ 
$\infty \ ,$ \qquad if $\ r>r_{\max }$%
\end{tabular}%
\right .
\end{equation}

In the region $r<r_{\max }$ the Schr\"{o}dinger equation becomes
\begin{equation}
-\frac{1}{2}\frac{d^{2}\tilde{u}_{nl}}{dr^{2}}\ + \ V_{\rm eff}(r)\ \tilde{u}_{nl}\ = \ E_{\max }\,
\tilde{u}_{nl}\, .
\label{eqrmax}
\end{equation}
%%in which the effective potential is given by:
%%
%%\begin{equation}
%%\label{Veffp}
%%V_{\rm eff}(r)=\left( \frac{l(l+1)}{2r^{2}}+w(r)-\frac{1}{r}\right) .
%%\end{equation}
This differential equation (\ref{eqrmax}) will be solved by 
the finite difference and finite element methods. The function 
$\tilde{u}$ satisfies the Dirichlet boundary conditions:
\begin{eqnarray}
\label{bouncond}
\tilde{u}_{nl}(0) \ = \ 0 \ , \\
\tilde{u}_{nl}(r_{\max }) \ = \  0 \ .  \nonumber
\end{eqnarray}
%i. e., the wave function $\tilde{u}$ %must fulfill Dirichlet boundary %conditions (DBC).
The original energy eigenvalues and eigenfunctions of the free (unbounded) 
system (\ref{scheq}) are recovered in the limit $r_{\max }\rightarrow \infty$,
\begin{equation}
\tilde{u}_{nl}(r) \, \rightarrow\, u_{nl}(r)\,,\qquad 
E_{\max }\, \rightarrow \, E \ .
\end{equation}

%%****
As was mentioned above, to find the solution of the
radial Schr\"{o}dinger equation~(\ref{eqrmax}) in the region $\left[ 0,r_{\max }\right] $,
the domain is splitted in $N$ subintervals of equal length $h$:
\begin{equation}
0=r_{1}<r_{2}<\cdot \cdot \cdot <r_{N+1}=r_{\max },
\end{equation}
where $h=r_{j+1}-r_{j}$.

Now, a second
order centred difference approximation to the second derivative can be used, namely
\begin{equation}
\tilde{u}^{^{\prime \prime }}\ = \ \frac{\tilde{u}(r_{j+1})-2\tilde{u}(r_{j})+%
\tilde{u}(r_{j-1})}{h^{2}} \ + \ O(h^{2})\ .
\end{equation}
Hence, the Schr\"{o}dinger equation can be written as an eigenvalue problem $H\,C\,=\,E\,C$
where the matrix $H$ is a tridiagonal matrix whose elements different of
zero are given by:
\begin{eqnarray}
H_{ij} &=&\frac{1}{h^{2}}+V_{\rm eff}(r_{i})\ , \\
H_{i,i+1} &=&-\frac{1}{2h^{2}}\ ,  \nonumber \\
H_{i,i-1} &=&-\frac{1}{2h^{2}} \ .  \nonumber
\end{eqnarray}
whereas the vector $C$ contains the values of $\tilde{u}$ evaluated on the grid $%
C_{i}=\tilde{u}(r_{i}).$

To improve the accuracy of the calculation one can use a forth order centred
difference approximation to the second derivative.

%--------------------------------------------------------------------------
%--------------------------------------------------------------------------
\subsection{Finite element method}

The finite element method (FEM) is a method based on the discretization of
the space in elements and the use of polynomial interpolating functions on each element.
This method is widely used in engineering, classical physics and quantum
mechanics problems, among others. The discretization is based on the
reformulation of the differential equation as an equivalent variational
problem. The Garlekin methods are employed in the corresponding minimization \cite{Pepper}. Usually, one can identify the following steps to
solve the associated differential equation: 
%$i$) the variational formulation of the problem,
$i)$ to present the problem in a variational formulation, 
$ii$) a discretization of the domain using FEM, and finally, $iii$) to find the solution of the discrete problem,
which may consist of the solution of a system of simultaneous equations or
an eigenvalue problem.

In Quantum Mechanics the FEM was used from a few year ago \cite{Kobus}, \cite{Friedman1}, 
\cite{Friedman2}, \cite{RMohan1}, \cite{RMohan2}, \cite{guimaraes2005study}, \cite{Moritz}. 
An excellent introduction of FEM in Quantum Mechanics is found in the Ram-Mohan's 
book~\cite{RMohan2}. Only the key points of the method will be presented here.
%Here we will only mention some general points of the method.

The time independent Schr\"{o}dinger equation for a particle of mass $m$
subjected to a potential energy $V({\mathbf r})$ is given by:
\begin{equation}
\label{TISE}
\bigg[\,-\frac{\hbar ^{2}}{2\,m}\Delta\ +\ V({\mathbf r})\,\bigg]\,\psi ({\mathbf r})\ =\ E\,\psi ({\mathbf r}) \ .
\end{equation}
This equation can be obtained as an extreme value of the following action
integral $I$:
\begin{equation}
I=\int d^{3}r\left[ \frac{\hbar ^{2}}{2m}\nabla \psi ^{\ast }\cdot \nabla
\psi +\psi ^{\ast }\left( \,V\,-\,E\,\right) \psi \right] \ ,
\end{equation}
where $\psi ^{\ast }$ and $\psi $ are considered as two independent
"fields". It is assumed that $\psi $ is continuous up to its second
derivative. By varying the action $I$ with respect to $\psi ^{\ast }$ we
obtain the Schr\"{o}dinger equation (\ref{TISE}). In the present problem only the dynamics of the $r$
coordinate is not trivial. The problem is completely analogous to the
one-dimensional case in the $r$-space, the variable $r$ varies in the interval $\left[
0,r_{\max }\right] $. Hence, the action integral $I$ (in atomic units) is
reduced to:
\begin{equation}
I=\int_{0}^{r_{\max }}dr\left\{ \frac{1}{2}\frac{d\psi ^{\ast }}{dr}\frac{%
d\psi }{dr}+\psi ^{\ast }\left[ V_{\rm eff}(r)-E\right] \psi \right\} \ ,
\label{IC}
\end{equation}
where $\psi=\psi(r)$, and $V_{\rm eff}(r)$ is an effective potential $V_{\rm eff}(r) = V(r) + \frac{l\,(l+1)}{2\,r^2}$, c.f.~(\ref{Veffp}).

Now, the interval $\left[ 0,r_{\max }\right] $ is divided into small subintervals
called \emph{elements}. The action integral (\ref{IC}) can be decomposed as the sum of the action computed in
each element,

\begin{equation}
I\ = \ \sum_{j=1}^{n}I^{(j)}\ ,
\end{equation}
where $n$ is the number of elements and $I^{(j)}$ is the action integral
evaluated on the $jth$ element.
Explicitly, the wave function $\psi
_{j}$ defined in the $jth$ element is expanded as a linear combination
\begin{equation}
\label{psijj}
\psi _{j}(r)\ = \ \sum_{j=1}^{n} c_{j}\,N_{j}(r)\ ,
\end{equation}
where $c_{j}$, $j=1,2,\ldots,n$, are unknown coefficients to be determined whereas $N_{j}(r)$ are interpolating
polynomials. These are defined for the $jth$ element and they are identically zero out of
this element.

The basic idea is to make the variation of the action integral with respect to
the coefficients $c_{i}^{\ast }$,
\begin{equation}
\frac{\delta I}{\delta c_{i}^{\ast }}\ =\ 0\ , \quad i=1,2,\ldots,n \ ,
\end{equation}
and by solving these equations to obtain the optimal energies and
eigenfunctions.

In particular, Guimaraes and Prudente \cite{guimaraes2005study} developed an alternative version of FEM called p-Finite Element Method (pFEM) to study the confined hydrogen atom, and Nascimento et. al. \cite{NPGM:2011} employed a pFEM version to study the electron structure of endohedrally confined atoms using an attractive gaussian potential to model atoms inside fullerenes.

As $r_{\max }$ becomes very large the energies of a confined system
approach those of the confinement-free system. In practice, a large value of both $r_{\max }$ and the parameter $n$, and a polynomial degree for the $N_{j}(r)$ appearing in (\ref{psijj}) are chosen and then, the generalized eigenvalue
problem is solved. For a fixed $r_{\rm max}$, by increasing either the value of $n$ or the polynomial degree, or both, a higher precision in the results can be achieved as we will explain in the next section.

\subsection{Artificial neural networks}

Artificial intelligence has emerged as a  collection of computational techniques which seek to mimic the human brain in order to complete tasks for which standard algorithms lead to partially satisfactory results or are costly to implement \cite{jackson2019introduction, you2020identification, lollie2022high, bhusal2022smart}. Particularly, neural networks are artificial intelligence algorithms inspired by the workings of neurons in the human brain. These algorithms have demonstrated the capacity of pinpointing relevant specific pieces of information `buried' in huge data sets and unveiling complex non-linear relationships between the inputs and target, which would be all but impossible to accomplish through a standard visual inspection \cite{murdoch2013inevitable, quiroz2021identification, villegas2022identification}. 

In this work, we implement a neural network to estimate the eigenvalues of the radial Schr\"odinger equation (\ref{scheq}) for different positions of the peak of the Gaussian, $r_c$. This neural network consists of a hidden layer and an output layer under a feed-forward architecture. In general, the output of each neuron before the activation function reads, 
\begin{equation}
z=\sum_{i=1}^N \omega_i x_i,     
\end{equation}
where $\omega_i$ are the synaptic weights, $x_i$ are the inputs, and $N$ is the number of inputs. Importantly, all the neurons in the output layer contain linear activation functions whereas the neurons in the hidden layer have sigmoid functions given by
\begin{equation}
    \alpha(z)=\frac{1}{1+e^{-z}}.
\end{equation}

Synaptic weights of the neural network are optimized with Levenberg-Marquardt backpropagation method in a direction that minimizes the mean squared error \cite{marquardt1963algorithm, hagan1994training}. This method approaches second-order training speed without having to compute the Hessian matrix. Because the performance function is given by a sum of squares then the Hessian matrix can be approximated as $\mathbf{H}=\mathbf{J}^T\mathbf{J}$, where $\mathbf{J}$ is the Jacobian matrix. Using this approximation, the synaptic weights can be updated by the following expression,
\begin{equation}
    \mathbf{\omega}_{k+1}=\mathbf{\omega}_{k}-[\mathbf{H}+\mu I]^{-1}\mathbf{J}^{T}\mathbf{e}
\end{equation}
where $k$ denotes the $k$th iteration, $\mu$ is the learning rate and $\mathbf{e}$ is the vector of networks errors. To train the neural network, we use a subset of the dataset that contains eigenvalues of the radial Schr\"odinger equation (\ref{scheq}) for different positions $r_c$ of the peak of the Gaussian. These eigenvalues are calculated by the Lagrange-mesh method. After the training stage, the neural network is able to predict the eigenvalues of the whole dataset with a coincidence in six significant digits at least.

%--------------------------------------------------------------------------------------
%--------------------------------------------------------------------------------------
\section{Results and discussion}
For the hydrogen atom in the presence of a Gaussian
confining spherical shell, the energies, eigenfunctions and expectation values of $r$ for the first
six states of $s, p$ and $d$ symmetries are accurately calculated. In order to be able to compare with previous results, the values of the parameters of the Gaussian potential (\ref{gaussV}) that we consider in detail are: $\omega_0=0.5$~a.u., 
$r_c= 2.5$~\r{A} and $3.54$~\r{A} and $\sigma= 0.26, 0.57$ and $1.59$ \r{A}. The corresponding results are shown 
in Tables \ref{Tabl0Rc354}, \ref{TPstates} and \ref{TDstates}. 

\begin{figure}
\subfloat[]{\includegraphics[scale=1.2]{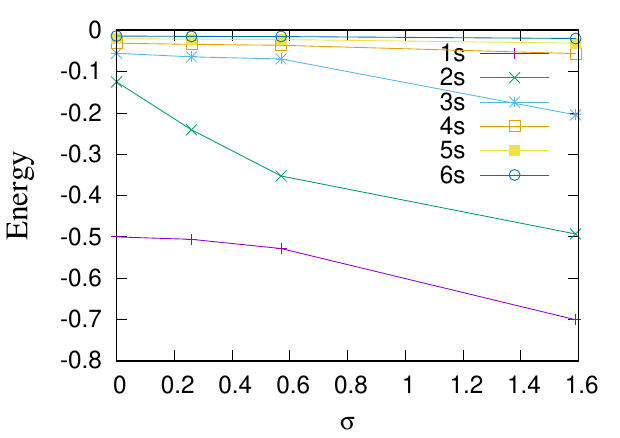}\label{fsa}} 
\subfloat[]{\includegraphics[scale=1.2]{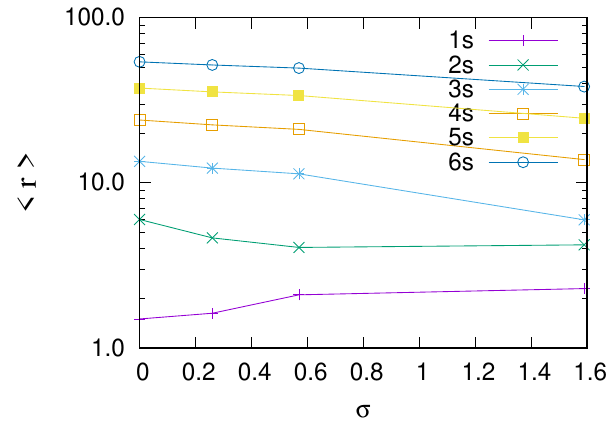}\label{fsb}}\\
\subfloat[]{\includegraphics[scale=1.2]{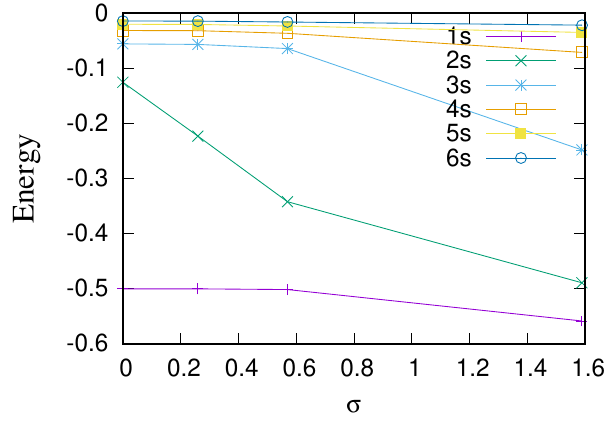}\label{fsc}} 
\subfloat[]{\includegraphics[scale=1.2]{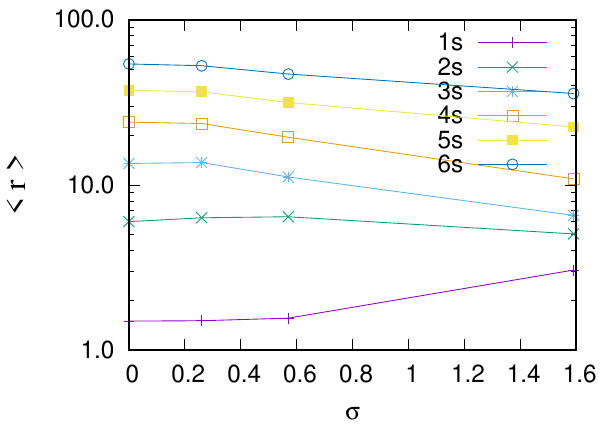}\label{fsd}}\\
\caption{Energy and expectation value for the s-states of the hydrogen atom as 
a function of the width $\sigma$ of the Gaussian potential~(\ref{Gpot})  with 
$\omega_0=0.5$ a.u. Figures (a) and (b) correspond to $r_c=2.50$~\r{A}  and  
(c) and (d) to $r_c=3.54$~\r{A}.}
\label{fsstates}
\end{figure}

Before discussing these results let us briefly mention some details about the Lagrange-mesh method.
In the system (\ref{lmmeqs}), there are two free parameters: the size of the mesh $N$
and the scaling factor $h$. The optimal values of this two parameters depend on the considered 
state of the system as well as on the value of the parameters occurring in the Gaussian potential. In general, the results presented in 
Tables \ref{Tabl0Rc354}, \ref{TPstates} and \ref{TDstates} for states $s$, $p$ and $d$, 
respectively, are obtained with a mesh of at least $N=250$ points and $h$ in some interval 
between $h\in [0.02-1.0]$. 
The convergence of the LMM is determined by the stability of the results with respect 
to an increase in the size of the base $N$ and the variation of $h$. 
Table \ref{Tabl0Rc354} presents the results for the $1s, \dots, 6s$ 
states. For $r_c=3.54$~\r{A} a comparison with~\cite{LH:2012} is possible for the levels $1s$, 
$2s$, $3s$, $4s$ with $\sigma = 1.59, 0.57$ and $0.26$~\r{A} where it 
can be seen a complete agreement in 7 decimal digits (for $1s$ and $2s$ states) and 
8 decimal digits (for $3s$ and $4s$). 
For completeness, the case $\sigma=0$ (the free hydrogen atom) is also presented.
For all these $s$-states, the energy decreases by increasing the value of $\sigma$ as can 
be seen in Figure~\ref{fsc}. On the other hand, the presence of the Gaussian potential 
has an important effect on the expectation value of $r$ as indicated in the seventh 
column of Table~\ref{Tabl0Rc354} (see also Figure~\ref{fsd}): $i$) as a function of 
$\sigma \in[0.00,1.59]$~\r{A}, $\langle r \rangle$ increases for the ground state, whilst $ii$) for the states $4s, \dots, 6s$, 
$\langle r \rangle$ decreases. Columns 3 and 4 of Table~\ref{Tabl0Rc354} 
(see also Figures~\ref{fsa} and \ref{fsb}) present the results of the energy and the 
expectation value $\langle r \rangle$ when the center of the Gaussian potential is 
$r_c=2.50$~\r{A}. As well as for $r_c=3.55$~\r{A}, the energy decreases as a function 
of $\sigma$. The behaviour of the expectation value $\langle r \rangle$ is depicted in Figure~\ref{fsb}. 
This effect on $\langle r \rangle$ reflects how the electronic charge is attracted 
to the Gaussian part of the potential.

\begin{figure}
\subfloat[]{\includegraphics[scale=1.2]{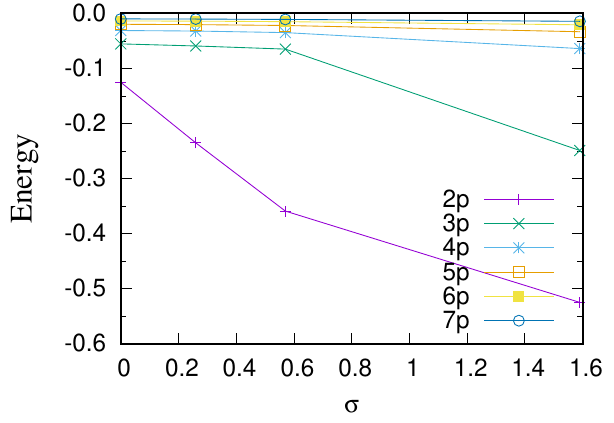}\label{fpa}} 
\subfloat[]{\includegraphics[scale=1.2]{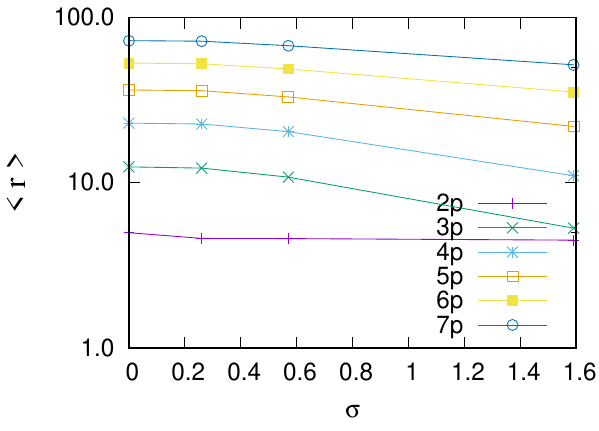}\label{fpb}}\\
\subfloat[]{\includegraphics[scale=1.2]{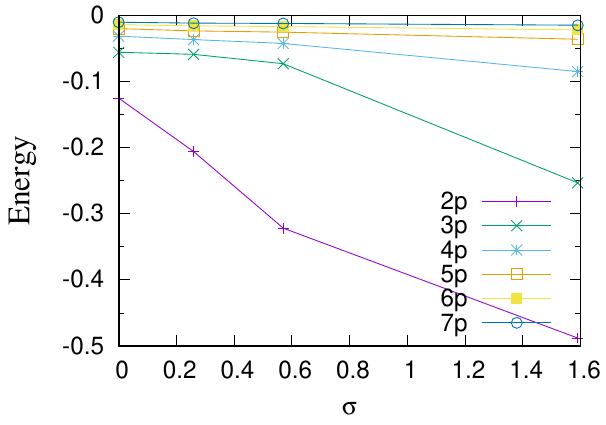}\label{fpc}} 
\subfloat[]{\includegraphics[scale=1.2]{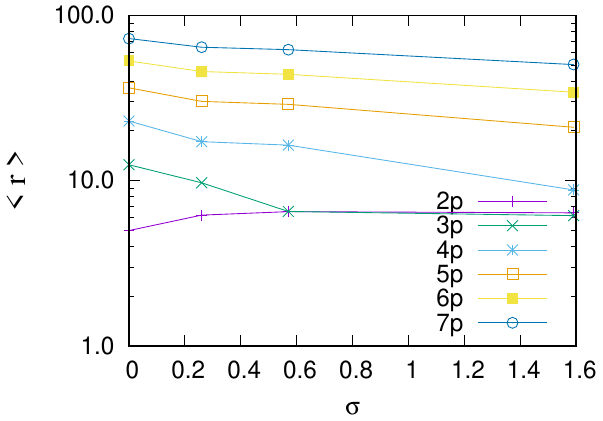}\label{fpd}}\\
\caption{Energy and expectation value for the p-states of the hydrogen atom as 
a function of the width $\sigma$ of the Gaussian potential~(\ref{Gpot})  with 
$\omega_0=0.5$ a.u. Figures (a) and (b) correspond to $r_c=2.50$~\r{A}  and  
(c) and (d) to $r_c=3.54$~\r{A}.}
\label{fsstates}
\end{figure}

States with $l=1$ ($2p, \dots, 7p$) are presented in Table~\ref{TPstates} for $r_c=2.50$ and 
$3.54$~\r{A}. In both cases, the system gets more bound as the value of $\sigma$ 
increases (see Figures~\ref{fpa} and \ref{fpc}). For $r_c=3.54$~\r{A} a comparison 
with~\cite{LH:2012} is also possible: we see a complete agreement in 7 decimal digits 
for the $2p$ state and 8 decimal digits for the $3p$ and $4p$ states for the three values 
of $\sigma = 0.26, 0.57$ and  $1.59$~\r{A}. 
The expectation value $\langle r \rangle$ is shown in Figures~\ref{fpb} and \ref{fpd}.

Results of the energy and the expectation value for $l=2$ ($3d, \dots, 8d$) are displayed 
in Table~\ref{TDstates} for $r_c=2.50$ and $3.54$~\r{A}. For both values of $r_c$,
by increasing $\sigma$ the energy becomes more negative. The expectation value 
$\langle r \rangle$ exhibits a decreasing behavior with the increasing of $\sigma$ for all 
states except the lowest $3d$ state, which decreases and eventually increases.  

The finite difference method was implemented at second and fourth order in Matlab. Calculations using second order finite differences require a very large number $n$ of nodal points, which implies to diagonalize very large matrices and therefore a significant computational time is involved. For this reason we decided to use fourth order finite differences, in which case the error in the solution of eigensystem is of order $h^4$, here $h$ being the distance between two consecutive nodal points. For $\omega_0=0.5$ a.u., $\sigma=0.26$~\r{A} and $r_c=3.54$~\r{A}, by using $h=0.01$, we obtained an accuracy of 9 or 10 decimal places in energy eigenvalues when they are compared with the results calculated with the Lagrange-mesh method and the finite element method (see below), respectively. For $l=0$ and $l=1$, $r_{\rm max}=160$, whereas for $l=2, r_{\rm max}=200$. Analogous results are found for other values of $\omega_0, \sigma$ and $r_{\rm max}$ and different values of $l$. {It is worth mentioning that by increasing the value of $r_{\rm max}$ and $n$ the agreement with the results of the LMM is in all digits.}

When the finite element method is applied, the agreement with the results of 
the LMM is in 8 decimal digits.
It is worth mentioning that the MATHEMATICA software package allows us to calculate the solutions of the eigenvalue problem (\ref{scheq}) as well. It can be easily done using the NDEigensystem command (based on the finite element method) which, in general, provides the smallest eigenvalues and eigenfunctions of the involved linear differential operator on a certain finite region. Therefore, again
is convenient to confine the system inside an impenetrable spherical cavity of radius $r_{max} \sim 150$. For the lowest states, in Table \ref{RMathe} we display the relative difference $\frac{|E_M - E_{LM}| }{|E_{LM}|}$ between the energy $E_M$ obtained with MATHEMATICA and the corresponding value $E_{LM}$ computed in the Lagrange-Mesh approach. The calculations were run in MATHEMATICA $12.3$ .

Finally, we train a regression neural network to estimate the energies of the 
hydrogen atom for the 1$s$ state as a function of the center of the inverted 
Gaussian potential (\ref{Gpot}), $r_c$. The training and testing data are 
generated by the Lagrange-Mesh method. After training, the neural network can 
predict the energy until six significant digits. Remarkably, our algorithm 
takes 40 $\mu$s to calculate the energy for a given value of $r_c$.
For $\omega_0=0.5$~a.u. and $\sigma=0.26$~\r{A}, Table~\ref{ia} 
displays the results obtained for the energy as a function of 
$r_c$ expressed as $r_c=\lambda\,r_0 $ where $r_0=3.54$~\r{A} 
and $\lambda$ is a factor specified in the first column. 

In each of these tables we compare the results obtained in the present work with those reported by Lin and Ho \cite{LH:2012} and, as can be seen, the methods presented in this work introduce an improvement to the energy values reported previously.

\begin{table*}[h]
\caption{Energy $E$ and expectation value $\langle r \rangle$ of the hydrogen atom of the $1s \dots 6s$ states in presence of a inverted Gaussian potential~(\ref{Gpot}) centered at $r_c=2.50$ and $3.54$~\r{A} with $w_0=0.5$~a.u. as a function of $\sigma$. The free case corresponds to $\sigma=0.0$. For $r_c=3.54$~\r{A} comparison is done with results presented in~\cite{LH:2012}.}
\centering
\begin{tabular}{l c| cc |ccc}
\hline
&&\multicolumn{2}{c|}{$r_c=2.50$~\r{A}}&\multicolumn{3}{c}{$r_c=3.54$~\r{A}}\\
\hline
$n$s&$\sigma$~\r{A}& $E$~a.u.&$\langle r \rangle$~a.u. & $E$~a.u.& \cite{LH:2012}& $\langle r \rangle$~a.u.\\
\hline
1s&0.00& -0.500000000000& 1.5000000000& -0.500000000000&            &  1.50000000000\\ 
  &0.26& -0.505803094144& 1.6250168416& -0.500226076582&  -0.5002261&  1.51021649828\\
  &0.57& -0.528322517980& 2.1023386012& -0.501274477556&  -0.5012745&  1.56269729613\\
  &1.59& -0.700338868882& 2.2929775088& -0.558460325443&  -0.5584603&  3.0539719446 \\
\hline                                			                             
2s&0.00& -0.125000000000& 6.0000000000& -0.125000000000&            &  6.00000000000\\
  &0.26& -0.240727300618& 4.6412959696& -0.222678640702&  -0.2226786&  6.33404911702\\
  &0.57& -0.352513053170& 4.0665509890& -0.341831613201&  -0.3418316&  6.42167789748\\
  &1.59& -0.493011698186& 4.2181317661& -0.489180987884&  -0.4891810&  5.0575771424 \\
\hline                                						     
3s&0.00& -0.055555555556& 13.500000000& -0.055555555556&            &  13.5000000000\\
  &0.26& -0.064455619315& 12.272530044& -0.056490840224& -0.05649084&  13.6508107402\\
  &0.57& -0.069257879934& 11.329276155& -0.063868006446& -0.06386801&  11.1726154458\\
  &1.59& -0.204794618028& 5.9698688054& -0.247983103635& -0.2479831 &   6.5426715036\\
\hline                                						     
4s&0.00& -0.031250000000& 24.000000000& -0.031250000000&            &  24.0000000000\\
  &0.26& -0.034143357000& 22.414917645& -0.031553410884& -0.03155341&  23.5646845778\\
  &0.57& -0.036157620392& 21.040808457& -0.036248303973& -0.03624830&  19.4517236193\\
  &1.59& -0.056183917194& 13.817270956& -0.070803944469& -0.07080395&  10.888054490 \\
\hline                                			                             
5s&0.00& -0.020000000000& 37.500000000& -0.020000000000&            &  37.5000000000\\
  &0.26& -0.021319794124& 35.527535710& -0.020260703774&            &  36.568599620 \\
  &0.57& -0.022355880943& 33.759026704& -0.022999812535&            &  31.5631588038\\
  &1.59& -0.030981035591& 24.577416804& -0.034734278117&            &  22.431757286 \\
\hline                                			                             
6s&0.00& -0.013888888889& 54.000000000& -0.013888888889&            &  54.0000000000\\
  &0.26& -0.014606403251& 51.633586493& -0.014097009471&            &  52.672305540 \\
  &0.57& -0.015206956729& 49.482686203& -0.015733494004&            &  46.8706368157\\
  &1.59& -0.019751915407& 38.280917640& -0.021384692368&            &  35.795142275 \\
\hline
\end{tabular}
\label{Tabl0Rc354}
\end{table*}

\begin{table*}[h]
\caption{Energy $E$ and expectation value $\langle r \rangle$ of the hydrogen atom of 
the $2p, \dots, 7p$ states in presence of a inverted Gaussian potential~(\ref{Gpot})
centered at $r_c=2.50$ and $3.54$~\r{A} with $w_0=0.5$~a.u. as a function of $\sigma$. 
The free case corresponds to $\sigma=0.0$. For $r_c=3.54$~\r{A} comparison is done with
results presented in~\cite{LH:2012}.}
\centering
\begin{tabular}{l c| cl | ccl}
\hline
&&\multicolumn{2}{c|}{$r_c=2.50$~\r{A}}&\multicolumn{3}{c}{$r_c=3.54$~\r{A}}\\
\hline
$n$s&$\sigma$~\r{A}& $E$~a.u.&$\langle r \rangle$~a.u. & $E$~a.u.& \cite{LH:2012}& $\langle r \rangle$~a.u.\\
\hline
2p& 0.00&  -0.125000000000& 5.0000000000& -0.125000000000&            & 5.0000000000\\
  & 0.26&  -0.235017656232& 4.6073217318& -0.205773905331& -0.2057739 & 6.1949410697\\
  & 0.57&  -0.358937721074& 4.6009805888& -0.321662542152& -0.3216625 & 6.4972955252\\
  & 1.59&  -0.524721262609& 4.4988472199& -0.487790459579& -0.4877905 & 6.4000682529\\
\hline                                  
3p& 0.00&  -0.055555555555& 12.500000000& -0.055555555555&            & 12.500000000\\
  & 0.26&  -0.059201294702& 12.295298412& -0.058921847548& -0.05892185& 9.7143674959\\
  & 0.57&  -0.064787981905& 10.822752863& -0.072949413182& -0.07294941& 6.5103348451\\
  & 1.59&  -0.248825019321& 5.3254757370& -0.253036633049& -0.25303663& 6.1503406694\\
\hline                                  
4p& 0.00&  -0.031250000000& 23.000000000& -0.031250000000&            & 23.000000000\\
  & 0.26&  -0.032157720549& 22.700493507& -0.036553597541& -0.03655360& 17.230927430\\
  & 0.57&  -0.034940841538& 20.400019294& -0.042230125318& -0.04223013& 16.420194663\\
  & 1.59&  -0.063945693111& 11.042617509& -0.084718765323& -0.08471877& 8.8066740094\\
\hline                                  
5p& 0.0 &  -0.020000000000& 36.500000000& -0.020000000000&            & 36.500000000\\
  & 0.26&  -0.020360792830& 36.096483134& -0.023349792551&            & 30.203357619\\
  & 0.57&  -0.021903580743& 33.091901133& -0.025258633309&            & 28.971093917\\
  & 1.59&  -0.033489056655& 21.958625107& -0.035932061579&            & 21.062981799\\
\hline                                  
6p& 0.00&  -0.013888888889& 53.000000000& -0.013888888889&            & 53.000000000\\
  & 0.26&  -0.014070782545& 52.498386898& -0.015883759185&            & 45.834836839\\
  & 0.57&  -0.014999966315& 48.833976319& -0.016767326295&            & 44.081940239\\
  & 1.59&  -0.020903914137& 35.390733125& -0.021816312170&            & 34.305718085\\
\hline                                  
7p& 0.00&  -0.010204081633& 72.50000000 & -0.010204081633&            & 72.500000000\\
  & 0.26&  -0.010309583975& 71.90456627 & -0.011458478666&            & 64.278040559\\
  & 0.57&  -0.010907611689& 67.59664010 & -0.011950076576&            & 62.104185020\\
  & 1.59&  -0.014343119913& 51.737977258& -0.014800187457&            & 50.435203259\\
\hline
\end{tabular}
\label{TPstates}
\end{table*}

\begin{table*}[h]
\caption{Energy $E$ and expectation value $\langle r \rangle$ of the hydrogen atom of 
the $3d, \dots, 8d$ states in presence of a inverted Gaussian potential~(\ref{Gpot})
centered at $r_c=2.50$ and $3.54$~\r{A} with $w_0=0.5$~a.u. as a function of $\sigma$. 
The free case corresponds to $\sigma=0.0$.}
\centering
\begin{tabular}{l c| cc | cc}
\hline
&&\multicolumn{2}{c|}{$r_c=2.50$~\r{A}}&\multicolumn{2}{c}{$r_c=3.54$~\r{A}}\\
\hline
$n$d&$\sigma$\r{A} & $E$~a.u.&$\langle r \rangle$~a.u. & $E$~a.u.&$\langle r \rangle$~a.u. \\
\hline
3d& 0.00&  -0.055555555555& 10.500000000& -0.055555555555& 10.500000000\\
  & 0.26&  -0.128471359504& 5.4492168184& -0.146974205177& 6.8949261986\\
  & 0.57&  -0.251571681618& 4.9862287243& -0.269235578841& 6.7272772114\\
  & 1.59&  -0.411853426848& 5.1273184744& -0.432795177458& 6.7292773865\\
\hline                                  
4d& 0.00&  -0.031250000000& 21.000000000& -0.031250000000& 21.000000000\\
  & 0.26&  -0.041811532197& 15.885456275& -0.038490481788& 17.852091651\\
  & 0.57&  -0.044189228465& 14.896944804& -0.040240022616& 16.883522700\\
  & 1.59&  -0.127836599792& 6.8812472011& -0.172635611475& 7.4407900868\\
\hline                                  
5d& 0.00&  -0.020000000000& 34.500000000& -0.020000000000& 34.500000000\\
  & 0.26&  -0.024569869745& 28.278727512& -0.022880682620& 30.857778724\\
  & 0.57&  -0.025611985197& 27.021828122& -0.023721575291& 29.575002055\\
  & 1.59&  -0.037461999111& 18.248819233& -0.039108470145& 17.049986326\\
\hline                                  
6d& 0.00&  -0.020000000000& 34.500000000& -0.013888888889& 51.000000000\\
  & 0.26&  -0.016327055430& 43.560989985& -0.015360981734& 46.717692906\\
  & 0.57&  -0.016886116520& 42.025765284& -0.015837354470& 45.132637791\\
  & 1.59&  -0.022489169307& 31.285420901& -0.023341455972& 29.854962236\\
\hline                                  
7d& 0.00&  -0.010204081633& 70.50000000 & -0.010204081633& 70.50000000 \\
  & 0.26&  -0.011667991818& 61.81202047 & -0.011065988156& 65.53681933 \\
  & 0.57&  -0.012004198432& 59.99767438 & -0.011362746680& 63.65745222 \\
  & 1.59&  -0.015168052460& 47.209127992& -0.015654825275& 45.536405971\\
\hline                                  
8d& 0.00&  -0.007812500000& 93.00000000 & -0.007812500000& 93.00000000 \\
  & 0.26&  -0.008763164361& 83.05090442 & -0.008363446047& 87.34012304 \\
  & 0.57&  -0.008981421027& 80.95756396 & -0.008560845234& 85.17084106 \\
  & 1.59&  -0.010955981002& 66.10195182 & -0.011258294889& 64.170188267\\
\hline
\end{tabular}
\label{TDstates}
\end{table*}

\begin{table*}[h]
\centering
\caption{Relative error $|E_M - E_{LM}|/|E_{LM}|$ as a function of the quantum numbers
$n$ and $l$ at $\omega_0=0.5$~a.u. Here $E_M$ is the energy obtained with MATHEMATICA
(finite element) whereas $E_{LM}$ corresponds to the result of Lagrange-Mesh method.
Results are given by the number followed by the power of 10.}
\begin{tabular}{c c| lll | lll}
\hline\hline
&&\multicolumn{3}{c|}{$r_c=3.54$~\r{A}}&\multicolumn{3}{c}{$r_c=2.5$~\r{A}}  \\
\hline
$n$&$\sigma$~\r{A} &  $l=0$  &  $l=1$ & $l=2$ &  $l=0$ &  $l=1$ &  $l=2$ \\
\hline
$1$& 1.59& $1.4(-10)$& $4.6(-10)$& $5.7(-10)$&  $4.9(-10)$& $2.6(-10)$& $4.4(-10)$\\
   & 0.57& $2.6(-10)$& $7.8(-10)$& $9.8(-10)$&  $2.5(-10)$& $3.9(-10)$& $6.4(-10)$\\
   & 0.26& $2.4(-10)$& \,\,\,\,\,$1(-9)$ & $1.7(-9)$ &  $3.8(-10)$& $6.6(-10)$& $1.4(-9)$ \\
\hline  
$2$& 1.59& $5.7(-10)$& $6.9(-10)$& \,\,\,\,\,$1(-9)$ &  $7.2(-10)$& $7.4(-10)$& $1.5(-9)$ \\
   & 0.57& $7.3(-10)$& $1.8(-9)$ & $3.5(-9)$ &  $4.5(-10)$& $2.8(-9)$ & $3.5(-9)$ \\
   & 0.26& \,\,\,\,\,$1(-9)$   & $2.4(-9)$ & $3.5(-9)$ &  $6.5(-10)$& $2.9(-9)$ & $3.6(-9)$ \\
  \hline  
$3$& 1.59& $5.5(-10)$& $1.8(-9)$ & $3.5(-9)$ &  $1.4(-9)$ & $2.6(-9)$ & $3.3(-9)$ \\
   & 0.57& $2.7(-9)$ & \,\,\,\,\,$3 (-9)$ & $2.5(-9)$ &  $2.7(-9)$ & $3.2(-9)$ & $2.7(-9)$ \\
   & 0.26& $2.9(-9)$ & \,\,\,\,\,$3 (-9)$ & $2.7(-9)$ &  $2.8(-9)$ & $2.9(-9)$ & $2.5(-9)$ \\
  \hline  
$4$& 1.59& $2.6(-9)$ & $2.9(-9)$ & $2.6(-9)$ &  $2.9(-9)$ & $2.9(-9)$ & $2.7(-9)$ \\
   & 0.57& $2.9(-9)$ & $2.5(-9)$ & $6.2(-9)$ &  $2.9(-9)$ & $3.6(-9)$ & $5.7(-9)$ \\
   & 0.26& $2.5(-9)$ & \,\,\,\,\,$3(-9)$ & $6.8(-9)$ &  $2.8(-9)$ & $4.3(-9)$ & \,\,\,\,\,$6(-9)$ \\
\hline\hline   
\end{tabular}
\label{RMathe}
\end{table*}

\begin{table*}[h]
%\centering
%\begin{minipage}[t]{1.8\linewidth}
%\scalebox{1.0}{
\caption{Energies of the hydrogen atom for the 1$s$ state as a function of the 
center of the inverted potential (\ref{Gpot}), $r_c$.   Here $E_{LM}$ is the 
energy obtained with the Lagrange-Mesh method and $E_{AI}$ is the energy with 
the trained neural network. These results were obtained using the following 
parameters: $\omega_0 = 0.5$ a.u., $\sigma = 0.4913287924027$ a.u.
and $r_c=\lambda\, r_0$ with $r_0 = 6.6896304811752$~a.u..}%\hfill
\begin{tabular}{c | c| c}
\hline
$\lambda$&$E_{LM}$& $E_{AI}$ \\
\hline
1/10000& -0.542088077914 & -0.542088671\\ 
1/5000 & -0.542202008934 & -0.542202015\\ 
1/1000 & -0.543121063114 & -0.543121329\\ 
1/500  & -0.544288919456 & -0.544288945\\ 
1/100  & -0.554393891488 & -0.554393308\\ 
1/60   & -0.563834032583 & -0.563834743\\ 
1/45   & -0.572373271619 & -0.572317326\\ 
1/35   & -0.582815298146 & -0.582815187\\ 
1/25   & -0.603110319312 & -0.603110065\\ 
1/22   & -0.613278713190 & -0.613278192\\ 
1/17   & -0.638659560757 & -0.638659364\\ 
1/10   & -0.705251046859 & -0.705251154\\ 
1/8.5  & -0.722411069471 & -0.722411243\\ 
1/7.5  & -0.731076180441 & -0.731076188\\ 
1/6    & -0.732174699082 & -0.732174236\\ 
1/5    & -0.717486035423 & -0.717486546\\ 
1/3.5  & -0.655478372832 & -0.655478355\\ 
1/2.5  & -0.581245674651 & -0.581245509\\ 
0.6    & -0.516590710000 & -0.516590888\\ 
0.7    & -0.506188282257 & -0.506188473\\ 
0.9    & -0.500704580760 & -0.500702969\\ 
1.1    & -0.500071154498 & -0.500074229\\ 
1.3    & -0.500006755585 & -0.500008335\\ 
1.5    & -0.500000614285 & -0.500000610\\ 
1.7    & -0.500000054036 & -0.500000563\\
\hline
\end{tabular}%}
\label{ia}
%\end{minipage}
\end{table*}

\clearpage

\section{Conclusions}
In summary, for the lowest states with angular momentum $l=0,1,2$ the energies and eigenfunctions of the hydrogen atom confined by a penetrable potential are presented. The confining barrier was modeled by an inverted Gaussian function $w(r)=  -\,\omega_0\, \textrm{exp}[-(r-r_c)^2/\sigma^2]$. The approximate solutions of the corresponding Schrödinger equation were determined by three different numerical methods: $i$) the Lagrange-mesh method, $ii$) the (fourth-order) finite difference and $iii$) the finite element method. As a complementary tool, we use an artificial neural network to interpolate/extrapolate the results.

Using the Lagrange-mesh method accurate energies with not less that 11 significant figures were obtained. The optimal values of the size of the mesh $N$ and the scaling factor $h$ depend on the state being studied as well as on the parameters of the confining potential $w(r)$. Since this method is not completely based on a variational principle, we must point out that the computed energies are not necessarily greater than or equal to the exact ones. However, in all known cases where the results are stable with respect to the variation of $N$ and $h$ it turns out that it converge rapidly and generates simple highly accurate solutions.

The finite difference method is a robust scheme and, similar to the LMM, 
easy to implement.
%The finite difference method is the most robust among the four methods employed in this work.
%Similar to the previous one it is easy to implement. 
The accuracy of the energy eigenvalues depends on the order of approximation of the kinetic energy operator, the state of the system, the number $n$ of nodal points, the radius $r_{\rm max}$ and the parameters of the potential. In the present work using a fourth-degree approximation for the kinetic energy operator we were able to obtain from 9 to 10 significant decimals. It should also be noted that this method, like the Lagrange-mesh method, is not based on the variational principle. 

In the case of the finite element method very precise energies (always from above the exact ones) can be calculated by simultaneously increasing $r_{\rm max}$, the number of elements in which the interval $[0,\,r_{\rm max}]$ is divided and the degree of the interpolating polynomials. Moreover, this method allows us to deal with problems in higher spatial dimensions with regular and irregular boundaries, which is not so easy to implement in the Lagrange-mesh and finite difference methods. 

Finally, the artificial neural network is computationally a faster efficient tool to compute the spectra. Nevertheless, for the training stage it requires to know in advance accurate results in several points on the space of parameters. In combination with the Lagrange-mesh or the finite difference method it significantly reduces the overall computational time, although with less accuracy. By means of the methods used in the present study, energies were obtained with higher accuracy than those reported in the literature.

\section*{Acknowledgements}

The authors are grateful to S. A. Cruz for their interest in the work and useful discussions. M.A.Q.-J. would like to thank the support from DGAPA-UNAM under Project UNAM-PAPIIT TA101023.  
%\appendix

\nocite{*}

\bibliography{apssamp}% Produces the bibliography via BibTeX.

\end{document}